\documentclass[conference]{IEEEtran}
\usepackage[top = 0.76 in, bottom = 1.04 in, left = 0.62in, right = 0.62 in]{geometry}
\IEEEoverridecommandlockouts
\usepackage{cite}
\usepackage{amsmath,amssymb,amsfonts}
\usepackage{algorithmic}
\usepackage{graphicx}
\usepackage{textcomp}
\usepackage{multirow}
\usepackage{xcolor}
\def\BibTeX{{\rm B\kern-.05em{\sc i\kern-.025em b}\kern-.08em
    T\kern-.1667em\lower.7ex\hbox{E}\kern-.125emX}}
\begin{document}

\title{Channel Modeling for Ultraviolet Non-Line-of-Sight Communications Incorporating an Obstacle}

\author{\IEEEauthorblockN{Tianfeng Wu\textsuperscript{1}, Fang Yang\textsuperscript{1}, Tian Cao\textsuperscript{2}, Ling Cheng\textsuperscript{3}, Yupeng Chen\textsuperscript{4},\\ Jian Song\textsuperscript{1,}\textsuperscript{5}, Julian Cheng\textsuperscript{6}, and Zhu Han\textsuperscript{7}}\\ 

\IEEEauthorblockA{\textsuperscript{1}Department of Electronic Engineering, Beijing National Research Center for Information Science and Technology, \\Tsinghua University, Beijing 100084, P. R. China \\ \textsuperscript{2}School of Telecommunications Engineering, Xidian University, Xi’an 710071, P. R. China\\\textsuperscript{3}School of Electrical and Information Engineering, University of the Witwatersrand, Johannesburg 2000, South Africa\\ \textsuperscript{4}School of Microelectronics, Southern University of Science and Technology, Shenzhen 518055, P. R. China\\ \textsuperscript{5}Shenzhen International Graduate School, Tsinghua University, Shenzhen 518055, P. R. China\\ \textsuperscript{6}School of Engineering, The University of British Columbia, Kelowna, BC, V1V 1V7, Canada\\ \textsuperscript{7}Department of Electrical and Computer Engineering, University of Houston, Houston, TX 77004 USA \\ Email: wtf22@mails.tsinghua.edu.cn, fangyang@tsinghua.edu.cn, caotian@xidian.edu.cn, ling.cheng@wits.ac.za, \\chenyp@sustech.edu.cn, jsong@tsinghua.edu.cn, julian.cheng@ubc.ca, hanzhu22@gmail.com}}

\maketitle

\begin{abstract}
Existing studies on ultraviolet (UV) non-line-of-sight (NLoS) channel modeling primarily focus on scenarios without any obstacle, which makes them unsuitable for small transceiver elevation angles in most cases. To address this issue, a UV NLoS channel model incorporating an obstacle was investigated in this paper, where the impacts of atmospheric scattering and obstacle reflection on UV signals were both taken into account. To validate the proposed model, we compared it to the related Monte-Carlo photon-tracing (MCPT) model that had been verified by outdoor experiments. Numerical results manifest that the path loss curves obtained by the proposed model agree well with those determined by the MCPT model, while its computation complexity is lower than that of the MCPT model. This work discloses that obstacle reflection can effectively reduce the channel path loss of UV NLoS communication systems.
\end{abstract}


\section{Introduction}
\IEEEPARstart{N}{owadays}, radio frequency (RF) technology is widely adopted in wireless communication scenarios due to its attractive benefits, for example, low cost and easy deployment. However, this technology also encounters certain issues such as spectrum scarcity, mutual interference, and communication security \cite{ref1,ref2}. To cope with these problems and promote the development of the next-generation wireless communication technology, much effort has been devoted to the investigation of millimeter-wave (mm-Wave), terahertz (THz) \cite{ref3}, infrared (IR) \cite{ref4}, visible light communications (VLC) \cite{ref5}, and ultraviolet (UV) communications \cite{ref6,ref7}. Compared to RF, UV~has benefits such as huge bandwidth without licensing, immunity to electromagnetic interference, and high confidentiality \cite{ref6}. In comparison with mm-Wave, THz, IR, and VLC, which mainly depend on line-of-sight (LoS) links for communications, UV can convey information via atmospheric scattering, i.e., non-line-of-sight (NLoS) links. Moreover, the background noise of the UV communication systems adopting 200-280 nm can be ignored. Since the ozone layer has a strong absorption impact on the solar radiation within this band \cite{ref8}. Motivated by these advantages, UV communications have great potential to enable future ubiquitous communications by integrating with existing wireless communication technologies. 

Although tremendous efforts have been implemented on UV NLoS communications, channel modeling is still an open issue owing to the complexity of application environments \cite{ref9,ref10}. From \cite{ref11,ref12}, it can be discovered that the existing research on UV NLoS channel modeling mainly focuses on scenarios without obstacles, which makes them unsuitable for the small transceiver elevation angles in most cases, e.g., scenarios with many buildings or mountainous areas. Since when transceiver elevation angles are small, obstacles will inevitably appear in their field-of-views (FoVs) and influence the propagation links of UV signals. As for the model in \cite{ref11}, it takes the obstacle's height and width into account and assumes that the thickness is infinitely large. Moreover, this model supposes that transceiver FoV axes are located on the same plane. Concerning the model in \cite{ref12}, it idealizes the obstacle as an infinite plane, which is perpendicular to the Z-axis. These constraints greatly limit the applicability of the two models. 

To address the aforementioned problems, we propose a UV NLoS channel model incorporating an obstacle in this paper, where the shape of the obstacle is modeled as the cuboid, and its thickness, width, height, and coordinates are considered to approach real communication environments. Considering that the received pulse energy is composed of two parts: the energy contributed by atmospheric scattering and the energy obtained from obstacle reflection, we derive them separately to facilitate understanding. For the former, a weighting factor is introduced to determine whether the propagation links of the UV photon are blocked or not. For the latter, the reflection pattern of the reflection surface is processed as the Phong model, which has been verified by outdoor experiments. Compared to the related models \cite{ref11,ref12}, which are derived based on the Monte-Carlo photon-tracing (MCPT) method, the proposed model is derived based on the integration method, causing its lower computation complexity and shorter calculation time.

The remainder of the paper is organized as follows. First,~the received pulse energy contributed by atmospheric scattering is derived in Section II. Then, the received pulse energy obtained from obstacle reflection is developed in Section III. Following that, the proposed model is verified and analyzed in Section~IV. Finally, conclusions are drawn in Section V.
\begin{figure}[t]  
\centering  
\includegraphics[scale=0.366]{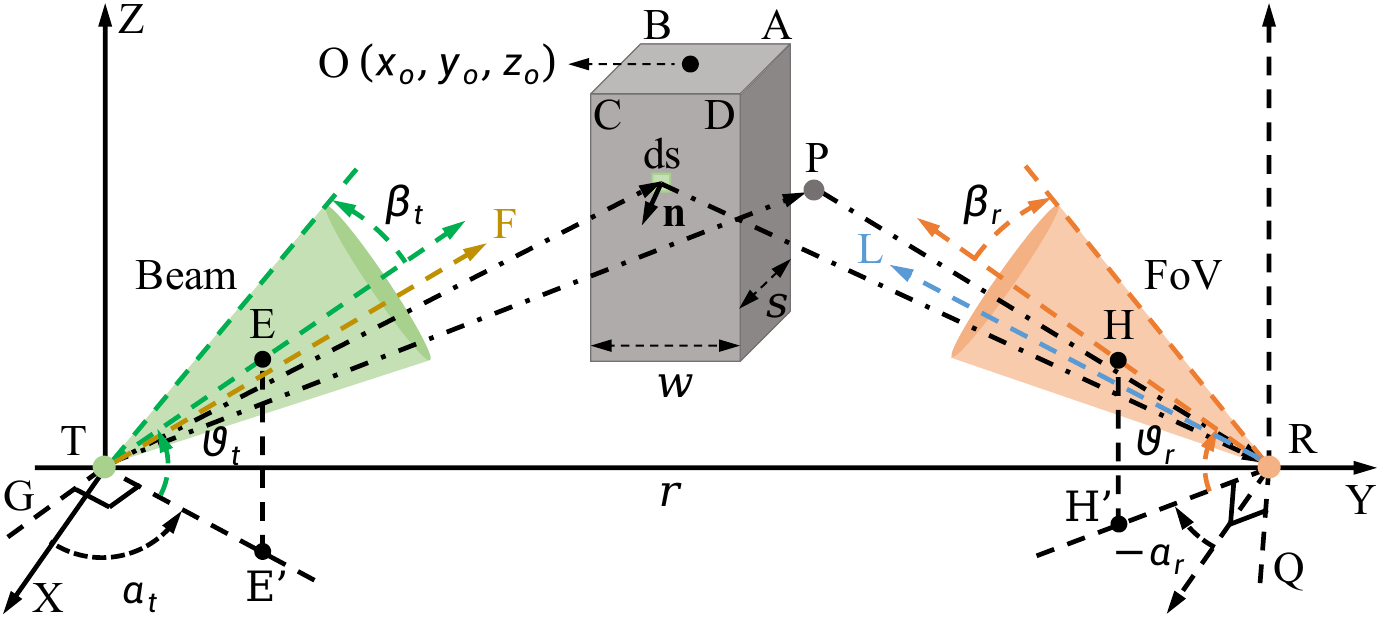}
\centering
\caption{Illustration of UV NLoS communication links with an obstacle.}
\label{Fig1}  
\end{figure}

\section{Derivation of Energy Contributed by Atmospheric Scattering} 
As presented in Fig.~\ref{Fig1}, the parameters of the system model are defined as follows: $\beta_t$ and $\beta_r$ represent the half-beam angle of the transmitter (T) and the half-FoV angle of the receiver (R), respectively; $\vartheta_t$ is the transmitter elevation angle, which is positive if taken anticlockwise from $\rm{TE'}$; $\vartheta_r$ is the receiver elevation angle and which is positive if taken clockwise from $\rm{RH'}$; $\alpha_t$ and $\alpha_r$ are the azimuth angles of the transmitter and receiver, respectively, which are positive if taken anticlockwise from X positive axis; $s$, $w$, and $\kappa$ are the obstacle's thickness, width, and height, respectively, while $\kappa$ refers to the height of the obstacle above XY; ${\rm{O}}\,(x_o, y_o, z_o)$ is the central position of the obstacle; $\tau$ and $\varepsilon$ are the distances from the scattering point P (or the reflection region d$\mathbb{U}$) to T and R, respectively; $r$ is the communication range; and $A_r$ is the detection area of the receiver aperture.

First, the parameter settings of the transceiver and obstacle are specified combined with real-life application environments. Given the symmetry of the obstacle, we set $x_o$ to $(-\infty,-s/2)$, $y_o$ to $(w/2,r-w/2)$, $\alpha_t$ to $[\pi/2,\pi)$, $\alpha_r$ to $(-\pi,-\pi/2]$, and $\vartheta_t$, $\vartheta_r$, $\beta_t$, and $\beta_r$ all to $(0,\pi/2)$, which satisfy the parameter setting requirements of UV NLoS systems for most scenarios. On this basis, the X and Y coordinates of points A, B, C, and D are given by: $x_a=x_b=x_o-s/2$, $y_a=y_d=y_o+w/2$, $y_b=y_c=y_o-w/2$, $x_c=x_d=x_o+s/2$, and $z_a=z_b=z_c=z_d=\kappa$. 

Second, planes $\mathcal{F}_{\vartheta}$ and $\mathcal{L}_{\sigma}$ are specified. As shown in Fig.~\ref{Fig1}, $\mathcal{F}_{\vartheta}$ represents the plane passing through TF and perpendicular to the plane $\rm{TEE'}$, and rotates around the line TG, where TG is located on the plane XY and perpendicular to $\rm{TE'}$, noting that the ray TF is located on the plane $\rm{TEE'}$ and the subscript $\vartheta$ represents the angle between $\boldsymbol{\rm{TE}}$ and $\boldsymbol{\rm{TF}}$, which is positive if taken anticlockwise from $\boldsymbol{\rm{TE}}$. $\mathcal{L}_{\sigma}$ denotes the plane passing through RL and perpendicular to $\rm{RHH'}$, and rotates around the line RQ, where RQ is located on the XY and perpendicular to $\rm{RH'}$. Note that here RL is located on the plane $\rm{RHH'}$, and the subscript $\sigma$ denotes the angle between $\boldsymbol{\rm{RH}}$ and $\boldsymbol{\rm{RL}}$, which is positive if taken clockwise from $\boldsymbol{\rm{RH}}$. Throughout the whole modeling process, $\delta_t=\vartheta_t+\vartheta$ and $\delta_r=\vartheta_r+\sigma$ are assumed to be greater than zero, and satisfy 
\begin{subequations}
\begin{equation}
\delta_t\le\min(\Theta_{t,a},\Theta_{t,b},\Theta_{t,c},\Theta_{t,d}),
\end{equation} 
\begin{equation}
\delta_r\le\min(\Theta_{r,a},\Theta_{r,b},\Theta_{r,c},\Theta_{r,d}),
\end{equation} 
\end{subequations}
where $\vartheta\in(-\beta_t,\beta_t)$ and $\sigma\in(-\beta_r,\beta_r)$, and $\Theta_{t,m}$ and $\Theta_{r,m}$ can be expressed as 
\begin{subequations} 
\begin{equation}
\Theta_{t,m}=\tan^{-1}\left(\frac{\kappa\sqrt{\cot^2{\alpha_t}+1}}{|x_m\cot{\alpha_t}+y_m|}\right),
\end{equation} 
\begin{equation}
\Theta_{r,m}=\tan^{-1}\left(\frac{\kappa\sqrt{\cot^2{\alpha_r}+1}}{|x_m\cot{\alpha_r}+y_m-r|}\right),
\end{equation} 
\end{subequations}
where the subscript $m$ can be $a$, $b$, $c$ or $d$, $\Theta_{t,m}$ is the angle between the plane $\mathcal{F}_{-\vartheta_t}$ and plane TGM, and $\Theta_{r,m}$ represents the angle between the plane $\mathcal{L}_{-\vartheta_r}$ and plane RQM.

When the pulse energy $\mathcal{Q}_t$ is emitted by the transmitter, the received energy contributed by atmospheric scattering, $\mathcal{Q}_{r,\rm{sca}}$, can be derived as \cite{ref13,ref14}
\begin{subequations}
\begin{equation}
\mathcal{Q}_{r,\rm{sca}}=\int_{\vartheta_{\min}}^{\vartheta_{\max}}\int_{\varpi_{\min}}^{\varpi_{\max}}\int_{\tau_{\min}}^{\tau_{\max}}\mathcal{K}\,\mathcal{G}_{\rm{wei}}\,\rm{d}\tau \rm{d}\varpi \rm{d}\vartheta,
\label{eq:3a}
\end{equation} 
\begin{equation}
\mathcal{K}=\frac{\mathcal{Q}_t\cos{\vartheta_v}{\rm{P}}(\cos{\vartheta_s})\exp[-k_e(\tau+\varepsilon)] A_r k_s |J_3|}{2\pi(1-\cos{\beta_t})\tau^2\varepsilon^2},
\label{eq:3b}
\end{equation}
\end{subequations} 
where $\varpi$ is the angle between $\boldsymbol{\rm{TF}}$ and $\boldsymbol{\rm{TF'}}$ ($\boldsymbol{\rm{TF'}}$ is located on the plane $\mathcal{F}_{\vartheta}$ and enclosed by the transmitter beam), which is positive when rotating anticlockwise from $\boldsymbol{\rm{TF}}$; $\vartheta_v$ is the angle between $\boldsymbol{\rm{RP}}$ and $\boldsymbol{\rm{RH}}$; $\vartheta_s$ is the angle between $\boldsymbol{\rm{TP}}$ and $\boldsymbol{\rm{PR}}$; ${\rm{P}}(\cos{\vartheta_s})$ is the scattering phase function, which is modeled as a weighted sum of the Rayleigh scattering phase function and the Mie scattering phase function \cite{ref15}; $k_e$ is the atmospheric extinction coefficient, which can be obtained by the addition of scattering coefficient $k_s$ and absorption coefficient $k_a$; and $\mathcal{G}_{\rm{wei}}$ is a weighting factor to determine whether the scattering point P in the overlap volume is valid or not. 

Suppose that $\rm{P}$ is located on the plane $\mathcal{F}_{\vartheta}$ and enclosed by the transmitter beam, its coordinates can be given by  
\begin{equation}
{\rm{P}:}
\begin{cases}
x=\tau\cos{\varpi}\cos{\delta_t}\,{\cos(\alpha_t+\phi)}{\sec{\phi}},\\
y=\tau\cos{\varpi}\cos{\delta_t}\,{\sin(\alpha_t+\phi)}{\sec{\phi}},\\
z=\tau\cos{\varpi}\sin{\delta_t},
\end{cases}
\label{eq:Pzb}  
\end{equation}
where $\phi=\tan^{-1}(\tan{\varpi}\sec{\delta_t})$. By transforming the differential volume from the cartesian coordinate system ${\rm{d}}x{\rm{d}}y{\rm{d}}z$ to the spherical coordinate system $|J_3|{\rm{d}}\tau{\rm{d}}\varpi{\rm{d}}\vartheta$, $|J_3|$ can be determined via Jacobian determinant and can be derived as
\begin{align}
|J_3| = & \,\frac{\partial x}{\partial \tau}\left(\frac{\partial y}{\partial \varpi}\frac{\partial z}{\partial \vartheta}-\frac{\partial y}{\partial \vartheta}\frac{\partial z}{\partial \varpi}\right)+\frac{\partial x}{\partial \varpi}\left(\frac{\partial y}{\partial \vartheta}\frac{\partial z}{\partial \tau}-\frac{\partial y}{\partial \tau}\frac{\partial z}{\partial \vartheta}\right)\nonumber\\
                    +& \,\frac{\partial x}{\partial \vartheta}\left(\frac{\partial y}{\partial \tau}\frac{\partial z}{\partial \varpi}-\frac{\partial y}{\partial \varpi}\frac{\partial z}{\partial \tau}\right)=\tau^2\cos{\varpi}.
\end{align} 

The unknowns in (\ref{eq:3a}) are the weighting factor and the upper and lower limits of the triple integral, i.e., $\vartheta_{\min}$, $\vartheta_{\max}$, $\varpi_{\min}$, $\varpi_{\max}$, $\tau_{\min}$, and $\tau_{\max}$. Based on the geometric relationships, $\vartheta_{\min}$ and $\vartheta_{\max}$ can be expressed as $-\beta_t$ and $\beta_t$, and $\varpi_{\min}$ and $\varpi_{\max}$ can be expressed as
\begin{equation}
\varpi_{\max}=\tan^{-1}\left(\frac{\sqrt{\tan^2{\beta_t}-\tan^2{\vartheta}}}{\sec{\vartheta}}\right)=-\varpi_{\min}. 
\end{equation}
As for $\tau_{\min}$ and $\tau_{\max}$, they are determined by the intersection situations between $\rm{TF'}$ and the receiver conical surface, which can be derived as
\begin{equation}
[\tau_{\min},\tau_{\max}]=
\begin{cases}
[\tau_0,+\infty), & \tau_0>0,\\ 
[\tau_2,+\infty), & \tau_1<0\,\,\rm{and}\,\,\tau_2>0,\\
[\tau_1,\tau_2], & \tau_1>0,\\
\emptyset, & \rm{otherwise},
\end{cases}
\end{equation}
where the real values $\tau_0$, $\tau_1$, and $\tau_2$ (by default, $\tau_2\ge\tau_1$) can be expressed as
\begin{equation}
\begin{cases}
\tau_0=-\xi_3/\xi_2, &\xi_1=0\,\,\rm{and}\,\,\xi_2\ne0,\\
\tau_{1,2}=\displaystyle{\frac{-\xi_2\pm\sqrt{\Delta}}{2\,\xi_1}}, &\xi_1\ne0\,\,\rm{and}\,\,\Delta\ge0,
\end{cases}
\end{equation} 
and $\Delta=\xi^2_2-4\,\xi_1\xi_3$. Here, $\xi_1$, $\xi_2$, and $\xi_3$ can be derived as
\begin{subequations} 
\begin{equation}
\begin{aligned}
\,\,\xi_1=&-[\cos{\vartheta_r}(\mathcal{A}_1\cos{\alpha_r}+\mathcal{A}_2\sin{\alpha_r})+\mathcal{A}_3\sin{\vartheta_r}]^2\\
                        &+(\mathcal{A}^2_1+\mathcal{A}^2_2+\mathcal{A}^2_3)\cos^2{\beta_r},
\end{aligned}
\end{equation}
\begin{equation}
\begin{aligned}
\xi_2=&\,2\,r[\cos{\vartheta_r}(\mathcal{A}_1\cos{\alpha_r}+\mathcal{A}_2\sin{\alpha_r})+\mathcal{A}_3\sin{\vartheta_r}]\\
                        &\times\cos{\vartheta_r}\sin{\alpha_r}-2\,r \mathcal{A}_2 \cos^2{\beta_r},
\end{aligned}
\end{equation}
\begin{equation}
\xi_3=r^2(\cos^2{\beta_r}-\cos^2{\vartheta_r}\sin^2{\alpha_r}),
\end{equation}
\end{subequations} 
where $\mathcal{A}_1$, $\mathcal{A}_2$, and $\mathcal{A}_3$ are the parametric equation coefficients of $x$, $y$, and $z$ in (\ref{eq:Pzb}), respectively, and $\tau$ is the corresponding parameter. Then, the value of $\mathcal{G}_{\rm{wei}}$ is investigated for different situations.

First, the intersection circumstances between the transmitter beam and the obstacle should be analyzed. Based on geometric relationships, the relationship among $\Omega_{t,\min}$, $\Omega_{t,\max}$, $\Psi_{t,\min}$, and $\Psi_{t,\max}$ are summarized in Table~\ref{Tab1}, where $\Omega_{t,\min}=\alpha_t+\varpi_{\min}-\pi/2$, $\Omega_{t,\rm{max}}=\alpha_t+\varpi_{\max}-\pi/2$, $\Psi_{t,\rm{min}}=\Psi_{t,dd'}$, and $\Psi_{t,\rm{max}}=\Psi_{t,bb'}$. $\Psi_{t,mm'}$ denotes the angle between $\boldsymbol{\rm{TK}}$ and $\boldsymbol{{\rm{TP}}_{t, m m'}}$, and can be given by
\begin{equation}
\Psi_{t,mm'}=\cos^{-1}\left(\frac{y_{mm'}\Xi_{t,c}-z_{t,mm'}\Xi_{t,b}}{\sqrt{\Xi^2_{t,b}+\Xi^2_{t,c}}||\boldsymbol{{\rm{TP}}_{t, mm'}}||}\right),
\end{equation}
where
\begin{subequations}
\begin{equation}
\Xi_{t,a}=-{\cos{\alpha_t}\sec^2{\vartheta}\sin(2\delta_t)\tan{\varphi_t}},
\end{equation}
\begin{equation}
\Xi_{t,b}=-{\sin{\alpha_t}\sec^2{\vartheta}\sin(2\delta_t)\tan{\varphi_t}},
\end{equation}
\begin{equation}
\Xi_{t,c}={2\sec^2{\vartheta}\cos^2\delta_t\tan{\varphi_t}}.
\end{equation}
\end{subequations}
and $\varphi_t=\tan^{-1}[\sec{\delta_t}(\tan^2\beta_t-\tan^2\vartheta)^{1/2}\cos{\vartheta}]$. Here, TK~is the intersection ray between planes YZ and $\mathcal{F}_{\vartheta}$, and ${\rm{P}}_{t, m m'}$ represents the intersection point of the plane $\mathcal{F}_{\vartheta}$ with the line $\rm{MM'}$ (e.g., ${\rm{P}}_{t, a a'}$ and line $\rm{AA'}$), whose Z coordinate can be expressed as           
\begin{equation}
z_{t,mm'}=-(\Xi_{t,a} x_m+\Xi_{t,b} y_m)/\Xi_{t,c}.
\end{equation} 
\linespread{1.8}
\begin{table}[t]
\centering
\linespread{1.0}
\caption{Relationship Among $\Omega_{t,\rm{min}}$, $\Omega_{t,\rm{max}}$, $\Psi_{t,\rm{min}}$, and $\Psi_{t,\rm{max}}$}
\label{Tab1}
\begin{tabular}{| c | c |}
\hline
\multirow{2}{*}{\textbf{All Cases}}&$\boldsymbol{\Omega_{t,\rm{min}}}$, $\boldsymbol{\Omega_{t,\rm{max}}}$, $\boldsymbol{\Psi_{t,\rm{min}}}$, \textbf{and} $\boldsymbol{\Psi_{t,\rm{max}}}$\\
{}&\textbf{in descending order}\\
\hline 
{Case 1}&{$\Omega_{t,\rm{max}}>\Omega_{t,\rm{min}}\ge\Psi_{t,\rm{max}}>\Psi_{t,\rm{min}}$}\\
\hline
{Case 2}&{$\Omega_{t,\rm{max}}>\Psi_{t,\rm{max}}>\Omega_{t,\rm{min}}\ge\Psi_{t,\rm{min}}$}\\
\hline
{Case 3}&{$\Omega_{t,\rm{max}}>\Psi_{t,\rm{max}}>\Psi_{t,\rm{min}}\ge\Omega_{t,\rm{min}}$}\\
\hline
{Case 4}&{$\Psi_{t,\rm{max}}\ge\Omega_{t,\rm{max}}>\Omega_{t,\rm{min}}\ge\Psi_{t,\rm{min}}$}\\
\hline
{Case 5}&{$\Psi_{t,\rm{max}}\ge\Omega_{t,\rm{max}}>\Psi_{t,\rm{min}}>\Omega_{t,\rm{min}}$}\\
\hline
{Case 6}&{$\Psi_{t,\rm{max}}>\Psi_{t,\rm{min}}\ge\Omega_{t,\rm{max}}>\Omega_{t,\rm{min}}$}\\
\hline
\end{tabular}
\end{table}
\linespread{1.0} 
Second, the intersection circumstances between the receiver FoV and the obstacle are presented in Table~\ref{Tab2}, where $\Omega_{r,\rm{min}}$ and $\Omega_{r,\rm{max}}$ can be derived as $\Omega_{r,\rm{min}}=-\alpha_r-\pi/2-\mathcal{C}$ and $\Omega_{r,\rm{max}}=-\alpha_r-\pi/2+\mathcal{C}$, and
\begin{equation}
\mathcal{C}=\tan^{-1}\left(\cos{\sigma}\sqrt{\tan^2{\beta_r}-\tan^2{\sigma}}\right).
\end{equation} 
Here, $\Psi_{r,\rm{min}}=\Psi_{r,cc'}$ and $\Psi_{r,\rm{max}}=\Psi_{r,aa'}$, where $\Psi_{r, m m'}$ is the angle between $\boldsymbol{\rm{RS}}$ and $\boldsymbol{{\rm{RP}}_{r, m m'}}$, and can be given by
\begin{equation}
\Psi_{r,mm'}=\cos^{-1}\left(\frac{r\Xi_{r,c}-y_{mm'}\Xi_{r,c}+z_{r,mm'}\Xi_{r,b}}{\sqrt{\Xi^2_{r,b}+\Xi^2_{r,c}}||\boldsymbol{{\rm{RP}}_{r, m m'}}||}\right),
\label{eq:Psir} 
\end{equation} 
where 
\begin{subequations} 
\begin{equation}
\Xi_{r,a}=-{\cos{\alpha_r}\sec^2{\sigma}\sin(2\delta_r)\tan{\varphi_r}},
\end{equation} 
\begin{equation}
\Xi_{r,b}=-{\sin{\alpha_r}\sec^2{\sigma}\sin(2\delta_r)\tan{\varphi_r}},
\end{equation} 
\begin{equation}
\Xi_{r,c}={2\sec^2{\sigma}\cos^2\delta_r\tan{\varphi_r}},
\end{equation} 
\end{subequations}
and $\varphi_r=\tan^{-1}[\sec{\delta_r}(\tan^2\beta_r-\tan^2\sigma)^{1/2}\cos{\sigma}]$. Here, RS is the intersection ray between the planes YZ and $\mathcal{L}_{\sigma}$, and ${\rm{P}}_{r, m m'}$ represents the intersection point of the plane $\mathcal{L}_{\sigma}$ with the line $\rm{MM'}$ (e.g., ${\rm{P}}_{r, a a'}$ and line $\rm{AA'}$), whose Z coordinate can be given by
\begin{equation}
z_{r,mm'}=(r\Xi_{r,b}-y_m \Xi_{r,b}-x_m \Xi_{r,a})/\Xi_{r,c}.
\end{equation}  

Since $\vartheta$ and $\varpi$ are known, the intersection situations of $\rm{TF'}$ with the receiver conical surface can be easily determined. If $[\tau_{\min},\tau_{\max}]=\emptyset$, $\mathcal{G}_{\rm{wei}}=0$, which means that the emitted ray $\rm{TF'}$ has no energy contribution to $\mathcal{Q}_{r,\rm{sca}}$. If $[\tau_{\min},\tau_{\max}]\neq\emptyset$, the~value of $\mathcal{G}_{\rm{wei}}$ will be investigated from \textbf{Case 1} to \textbf{Case 5}. For \textbf{Case 6} and {\em{Condition 6}}, the value of $\mathcal{G}_{\rm{wei}}$ is always equal to $1$, since the propagation links of the UV signal are not blocked by the obstacle. For clarity, we only provide the circumstances where the value of $\mathcal{G}_{\rm{wei}}$ is equal to $1$ in the following derivation, and $\mathcal{G}_{\rm{wei}}=0$ in other circumstances. 
\linespread{1.8}
\begin{table}[t]
\centering
\linespread{1.0}
\caption{Relationship Among $\Omega_{r,\rm{min}}$, $\Omega_{r,\rm{max}}$, $\Psi_{r,\rm{min}}$, and $\Psi_{r,\rm{max}}$}
\label{Tab2}
\begin{tabular}{| c | c |}
\hline
\multirow{2}{*}{\textbf{All Conditions}}&$\boldsymbol{\Omega_{r,\rm{min}}}$, $\boldsymbol{\Omega_{r,\rm{max}}}$, $\boldsymbol{\Psi_{r,\rm{min}}}$, \textbf{and} $\boldsymbol{\Psi_{r,\rm{max}}}$\\
{}&\textbf{in descending order}\\
\hline 
{Condition 1}&{$\Omega_{r,\rm{max}}>\Omega_{r,\rm{min}}\ge\Psi_{r,\rm{max}}>\Psi_{r,\rm{min}}$}\\
\hline
{Condition 2}&{$\Omega_{r,\rm{max}}>\Psi_{r,\rm{max}}>\Omega_{r,\rm{min}}\ge\Psi_{r,\rm{min}}$}\\
\hline
{Condition 3}&{$\Omega_{r,\rm{max}}>\Psi_{r,\rm{max}}>\Psi_{r,\rm{min}}>\Omega_{r,\rm{min}}$}\\
\hline
{Condition 4}&{$\Psi_{r,\rm{max}}\ge\Omega_{r,\rm{max}}>\Omega_{r,\rm{min}}\ge\Psi_{r,\rm{min}}$}\\
\hline
{Condition 5}&{$\Psi_{r,\rm{max}}\ge\Omega_{r,\rm{max}}>\Psi_{r,\rm{min}}>\Omega_{r,\rm{min}}$}\\
\hline
{Condition 6}&{$\Psi_{r,\rm{max}}>\Psi_{r,\rm{min}}\ge\Omega_{r,\rm{max}}>\Omega_{r,\rm{min}}$}\\
\hline
\end{tabular}
\end{table}
\linespread{1.0} 

\textbf{Case 1:} $\Omega_{t,\rm{max}}>\Omega_{t,\rm{min}}\ge\Psi_{t,\rm{max}}>\Psi_{t,\rm{min}}$

In this case, $\mathcal{G}_{\rm{wei}}$ is always equal to $1$ for {\em{Condition 1}}. For {\em{Condition 2}}, $\Psi_{r,\rm{esp}}\in(\Psi_{r,\rm{max}},\Omega_{r,\rm{max}}]$, where $\Psi_{r,\rm{esp}}$ denotes the angle between $\boldsymbol{\rm{RS}}$ and $\boldsymbol{\rm{RP}}$, while for {\em{Condition 3}}, $\Psi_{r,\rm{esp}}$ $\in[\Omega_{r,\rm{min}},\Psi_{r,\rm{min}})\cup(\Psi_{r,\rm{max}},\Omega_{r,\rm{max}}]$. Further, $\Psi_{r,\rm{esp}}$ belongs to the interval $[\Omega_{r,\rm{min}},\Psi_{r,\rm{min}})$ for {\em{Condition 5}}.  

\textbf{Case 2:} $\Omega_{t,\rm{max}}>\Psi_{t,\rm{max}}>\Omega_{t,\rm{min}}\ge\Psi_{t,\rm{min}}$

In this case, the situations where $\mathcal{G}_{\rm{wei}}$ is always equal to one are provided first. For {\em{Condition 1}}, $\Psi_{t,\rm{esp}}\in(\Psi_{t,\rm{max}},\Omega_{t,\rm{max}}]$, where $\Psi_{t,\rm{esp}}$ is the angle between $\boldsymbol{\rm{TK}}$ and $\boldsymbol{\rm{TP}}$, while for {\em{Conditions 2}} and {\em{3}}, $\Psi_{t,\rm{esp}}$ belongs to $(\Psi_{t,\rm{max}},\Omega_{t,\rm{max}}]$ and $\Psi_{r,\rm{esp}}$ belongs to $(\Psi_{r,\rm{max}},\Omega_{r,\rm{max}}]$. Besides, the situation where $\Psi_{r,\rm{esp}}\in[\Omega_{r,\rm{min}},\Psi_{r,\rm{min}})$ needs to be incorporated for {\em{Condition 3}}, and this interval also applies to {\em{Condition 5}}.

Then, all possible situations where $\mathcal{G}_{\rm{wei}}$ is equal to one are provided. For {\em{Conditions 2}} to {\em{5}}, $\Psi_{t,\rm{esp}}\in[\Omega_{t,\min},\Psi_{t,cc'}]^{(\ref{eq:TPCD})}$, where the superscript (\ref{eq:TPCD}) denotes the equation number of the $||\boldsymbol{\rm{TP}}||$ constraint imposed on the related interval 
\begin{equation}  
||\boldsymbol{\rm{TP}}||<\frac{||\boldsymbol{{\rm{TP}}_{t,dd'}}||\sin{\psi_{t,dd'}}}{\sin(\Psi_{t,\rm{esp}}+\psi_{t,dd'}-\Psi_{t,dd'})}, 
\label{eq:TPCD}
\end{equation}
and $\Psi_{r,\rm{esp}}\in[\Omega_{r,\min},\Psi_{r,dd'}]^{(\ref{eq:RPCD})}$, $[\Psi_{r,\min},\Psi_{r,dd'}]^{(\ref{eq:RPCD})}$, $[\Omega_{r,\min},$ $\min(\Omega_{r,\max},\Psi_{r,dd'})]^{(\ref{eq:RPCD})}$, and $[\Psi_{r,\min},\min(\Omega_{r,\max},\Psi_{r,dd'})]$ $^{(\ref{eq:RPCD})}$, respectively, for {\em{Conditions 2}} to {\em{5}}, where the superscript (\ref{eq:RPCD}) denotes the $||\boldsymbol{\rm{RP}}||$ constraint imposed on the associated interval   
\begin{equation}  
||\boldsymbol{\rm{RP}}||<\frac{||\boldsymbol{{\rm{RP}}_{r,cc'}}||\sin{\psi_{r,cc'}}}{\sin(\Psi_{r,\rm{esp}}+\psi_{r,cc'}-\Psi_{r,cc'})}.
\label{eq:RPCD}
\end{equation}

\textbf{Case 3}: $\Omega_{t,\rm{max}}>\Psi_{t,\rm{max}}>\Psi_{t,\rm{min}}\ge\Omega_{t,\rm{min}}$

In this case, the circumstances where $\mathcal{G}_{\rm{wei}}$ is always equal to $1$ are provided first. For {\em{Condition~1}}, $\Psi_{t,\rm{esp}}\in[\Omega_{t,\min},\Psi_{t,\min})$ $\cup\,(\Psi_{t,\max},\Omega_{t,\max}]$, while for {\em{Conditions~2}}, {\em{3}}, {\em{4}}, and {\em{5}}, $\Psi_{t,\rm{esp}}$ $\in[\Omega_{t,\min},\Psi_{t,\min})$. Additionally, the situation where $\Psi_{t,\rm{esp}}\in$ $(\Psi_{t,\max},\Omega_{t,\max}]$ and $\Psi_{r,\rm{esp}}\in(\Psi_{r,\max},\Omega_{r,\max}]$ for {\em{Condition 2}} needs to be considered, while for {\em{Condition 5}}, the situation where $\Psi_{t,\rm{esp}}\in[\Psi_{t,\min},\Omega_{t,\max}]$ and $\Psi_{r,\rm{esp}}\in[\Omega_{r,\min},\Psi_{r,\min})$ must be incorporated, and for {\em{Condition 3}}, the following parts: i) $\Psi_{r,\rm{esp}}\in[\Omega_{r,\min},\Psi_{r,\min})$ and $\Psi_{t,\rm{esp}}\in[\Psi_{t,\min},\Psi_{t,\max}]$ and ii) $\Psi_{r,\rm{esp}}\in[\Omega_{r,\min},\Psi_{r,\min})\cup(\Psi_{r,\max},\Omega_{r,\max}]$ and $\Psi_{t,\rm{esp}}\in$ $(\Psi_{t,\max},\Omega_{t,\max}]$ must be taken into account. For the possible circumstances where $\mathcal{G}_{\rm{wei}}=1$, they are consistent with those of \textbf{Case 2}, where the item ``$\Omega_{t,\min}$'' of the corresponding $\Psi_{t,\rm{esp}}$ intervals requires to be changed to ``$\Psi_{t,\min}$''. 

\textbf{Case 4}: $\Psi_{t,\rm{max}}\ge\Omega_{t,\rm{max}}>\Omega_{t,\rm{min}}\ge\Psi_{t,\rm{min}}$

When $\Psi_{r,\rm{esp}}\in[\Omega_{r,\min},\Psi_{r,\min})$ in this case, $\mathcal{G}_{\rm{wei}}$ is always equal to $1$ for {\em{Conditions~3}} and {\em{5}}. Following that, the possible circumstances where $\mathcal{G}_{\rm{wei}}=1$ are presented. For {\em{Conditions 2}} to {\em{5}}, $\Psi_{t,\rm{esp}}\in[\Omega_{t,\min},\min(\Omega_{t,\max},\Psi_{t,cc'})]^{(\ref{eq:TPCD})}$, and the related $\Psi_{r,\rm{esp}}$ intervals are consistent with those of \textbf{Case 2}.

\textbf{Case 5}: $\Psi_{t,\rm{max}}\ge\Omega_{t,\rm{max}}>\Psi_{t,\rm{min}}>\Omega_{t,\rm{min}}$

In this case, the situations where $\mathcal{G}_{\rm{wei}}$ is always equal to $1$ are summarized first. When $\Psi_{t,\rm{esp}}\in[\Omega_{t,\min},\Psi_{t,\min})$, $\mathcal{G}_{\rm{wei}}=$ $1$ for {\em{Conditions~1}} to {\em{5}}, and when $\Psi_{t,\rm{esp}}\in[\Psi_{t,\min},\Omega_{t,\max}]$ and $\Psi_{r,\rm{esp}}\in[\Omega_{r,\min},\Psi_{r,\min})$, $\mathcal{G}_{\rm{wei}}=1$ for {\em{Conditions~3}} and {\em{5}}. Second, the possible circumstances where $\mathcal{G}_{\rm{wei}}$ is equal to $1$ are supplemented. For {\em{Conditions 2}} to {\em{5}}, $\Psi_{t,\rm{esp}}\in[\Psi_{t,\min},$ $\min(\Omega_{t,\max},\Psi_{t,cc'})]^{(\ref{eq:TPCD})}$, and the relevant $\Psi_{r,\rm{esp}}$ intervals are consistent with those of \textbf{Case 2}. 

Note that in the above analysis, we did not incorporate the energy contribution of the beam ray $\mathbb{L}^{t}_{\pm}$ and FoV rays $\mathbb{L}^{r}_\pm$ to the received energy $\mathcal{Q}_{r,\rm{sca}}$. Because in these scenarios, $\Omega_{t,\max}$ $=\Omega_{t,\min}$ or $\Omega_{r,\max}=\Omega_{r,\min}$, where $\mathbb{L}^{t}_{\pm}$ is the intersection ray between the plane $\mathcal{F}_{\pm\beta_t}$ and the transmitter beam, and $\mathbb{L}^{r}_\pm$ represent the intersection rays between planes $\mathcal{L}_{\pm\beta_r}$ and the receiver FoV. Their contribution to $\mathcal{Q}_{r,\rm{sca}}$ can be determined easily by referring to the above modeling process. For tractable analysis, the equations of $\mathbb{L}^{t}_{\pm}$ and $\mathbb{L}^{r}_\pm$ are provided as
\begin{align}
\mathbb{L}^{t}_{\pm}: &
\begin{cases}
x=\sec{\beta_t}\cos\delta_{t,\pm}\cos{\alpha_t}\,\omega,\\
y=\sec{\beta_t}\cos\delta_{t,\pm}\sin{\alpha_t}\,\omega,\\
z=\sec{\beta_t}\sin\delta_{t,\pm}\,\omega,
\end{cases}\\
\mathbb{L}^{r}_\pm: &
\begin{cases}
x=\sec{\beta_r}\cos\delta_{r,\pm}\cos{\alpha_r}\,\omega,\\
y=\sec{\beta_r}\cos\delta_{r,\pm}\sin{\alpha_r}\,\omega+r,\\
z=\sec{\beta_r}\sin\delta_{r,\pm}\,\omega,
\end{cases}
\end{align}
where $\delta_{t,-}=\vartheta_t-\beta_t$, $\delta_{t,+}=\vartheta_t+\beta_t$, $\delta_{r,-}=\vartheta_r-\beta_r$, and $\delta_{r,+}=\vartheta_r+\beta_r$, and the equation of the receiver conical surface can be given by
\begin{equation} 
\cos^2{\beta_r}=\frac{(\mathcal{N}_{r,x} x+\mathcal{N}_{r,y} y+\mathcal{N}_{r,z} z-\mathcal{N}_{r,y} r)^2}{x^2+(y-r)^2+z^2},     
\end{equation}
where $\mathcal{N}_{r,x}=\cos{\vartheta_r}\cos{\alpha_r}$, $\mathcal{N}_{r,y}=\cos{\vartheta_r}\sin{\alpha_r}$, and $\mathcal{N}_{r,z}=\sin{\vartheta_r}$.

\section{Derivation of Energy Attained from\\ Obstacle Reflection}
From Fig.~\ref{Fig1} it can be discovered that there is one reflection surface $\rm{CDD'C'}$, where $\rm{C'}$ and $\rm{D'}$ are the projection points~of C and D on the plane XY. Based on geometric relationships, the effective reflection region $\mathbb{U}_{\rm{cdd'c'}}$ can be expressed as
\begin{equation} 
\mathbb{U}_{\rm{cdd'c'}}:
\begin{cases}
x=x_c,\\
y_c \le y \le y_d,\\
\beta_{t,\rm{esp}} \le \beta_t,\\
\beta_{r,\rm{esp}} \le \beta_r,\\
z \le \kappa,
\end{cases}
\end{equation}
where $\beta_{t,\rm{esp}}$ and $\beta_{r,\rm{esp}}$ can be given by 
\begin{subequations}
\begin{equation} 
\cos \beta_{t,\rm{esp}}=(\mathcal{N}_{t,x}\,x+\mathcal{N}_{t,y}\,y+\mathcal{N}_{t,z}\,z)/\tau,  
\end{equation} 
\begin{equation} 
\cos \beta_{r,\rm{esp}}=[\mathcal{N}_{r,x}\,x+\mathcal{N}_{r,y}\,(y-r)+\mathcal{N}_{r,z}\,z]/\varepsilon,   
\end{equation}   
\end{subequations} 
and $\mathcal{N}_{t,x}=\cos{\vartheta_t}\cos{\alpha_t}$, $\mathcal{N}_{t,y}=\cos{\vartheta_t}\sin{\alpha_t}$, and $\mathcal{N}_{t,z}=\sin{\vartheta_t}$.   

Second, the energy contribution of the reflection region to the received energy $\mathcal{Q}_{r,\rm{ref}}$ is developed. When a pulse energy $\mathcal{Q}_t$ is transmitted by the transmitter, the unextinguished energy arriving at the differential reflection region d$\mathbb{U}$ can be derived as
\begin{equation} 
\mathcal{Q}_{{\rm{ref}},{\rm{d}}\mathbb{U}}=\frac{\mathcal{Q}_t \cos{\omega_i}\exp(-k_e \tau)}{2\pi(1-\cos{\beta_t})\tau^2}{\rm{d}}\mathbb{U},
\label{eq:dU}
\end{equation} 
where $\omega_i$ is the incidence angle of the UV photon, which can be expressed as $\cos{\omega_i}=-\boldsymbol{n \tau}^{\rm{T}}/\tau$, and $\boldsymbol{n}=[1,0,0]$ denotes the normal vector of $\mathbb{U}_{\rm{cdd'c'}}$. After the photons are reflected by ${\rm{d}}\mathbb{U}$, they arrive at the receiver aperture. Therefore, the received energy contributed by ${\rm{d}}\mathbb{U}$ can be expressed as
\begin{equation} 
{\rm{d}}\mathcal{Q}_{r,\rm{ref}}= r_r {\rm{I}}_r(\vartheta_1,\vartheta_2) \frac{A_r\cos{\vartheta_v}}{\varepsilon^{2}}\exp(-k_e\varepsilon) \mathcal{Q}_{{\rm{ref}},{\rm{d}}\mathbb{U}},
\label{eq:U}
\end{equation} 
where $r_r$ is the reflection coefficient, $\vartheta_1$ is the angle between $\boldsymbol{\varepsilon}$ and $\boldsymbol{n}$, $\vartheta_2$ is the angle between $\boldsymbol{\varepsilon}$ and $\boldsymbol{v_s}$, and $\boldsymbol{v_s}$ represents the direction vector of the specular reflection of $\boldsymbol{\tau}$. ${\rm{I}}_r(\vartheta_1,\vartheta_2)$ can be expressed as \cite{ref16}
\begin{equation}  
{\rm{I}}_r (\vartheta_1,\vartheta_2)=\eta\frac{\cos{\vartheta_1}}{\pi}+(1-\eta)\frac{m_s+1}{2\pi}\cos^{m_s}\vartheta_2,   
\end{equation}                
where $\eta$ is the percentage of the incident signal that is reflected diffusely and assumes values between $0$ and $1$, and $m_s$ is the directivity of the specular components. 

Substituting (\ref{eq:dU}) into (\ref{eq:U}) and integrating ${\rm{d}}\mathcal{Q}_{r,\rm{ref}}$ over the whole reflection region, the total received reflection energy can be derived as
\begin{subequations}
\begin{equation}
\mathcal{Q}_{r,\rm{ref}} = \iint_{\mathbb{U}_{\rm{cdd'c'}}} r_r \mathcal{Q}_t {A_r} \mathcal{Z}\,{\rm{d}}y{\rm{d}}z,
\end{equation}
\begin{equation}  
\mathcal{Z}=\frac{{\rm{I}}_r (\vartheta_1,\vartheta_2) \cos{\vartheta_v} \cos{\omega_i} \exp[-k_e(\tau+\varepsilon)]}{2\pi(1-\cos{\beta_t})\tau^2 \varepsilon^2}.
\end{equation} 
\end{subequations}     

By summing the scattered energy derived in Section II and the reflected energy obtained in Section III, the total received pulse energy can be given by
\begin{equation}
\mathcal{Q}_r=\mathcal{Q}_{r,\rm{sca}}+\mathcal{Q}_{r,\rm{ref}},
\end{equation} 
and the channel path loss for UV NLoS scenarios considering an obstacle can be expressed as $10\log_{10}(\mathcal{Q}_t/\mathcal{Q}_r)$.    
\begin{figure}[t]  
\centering  
\includegraphics[scale=0.36]{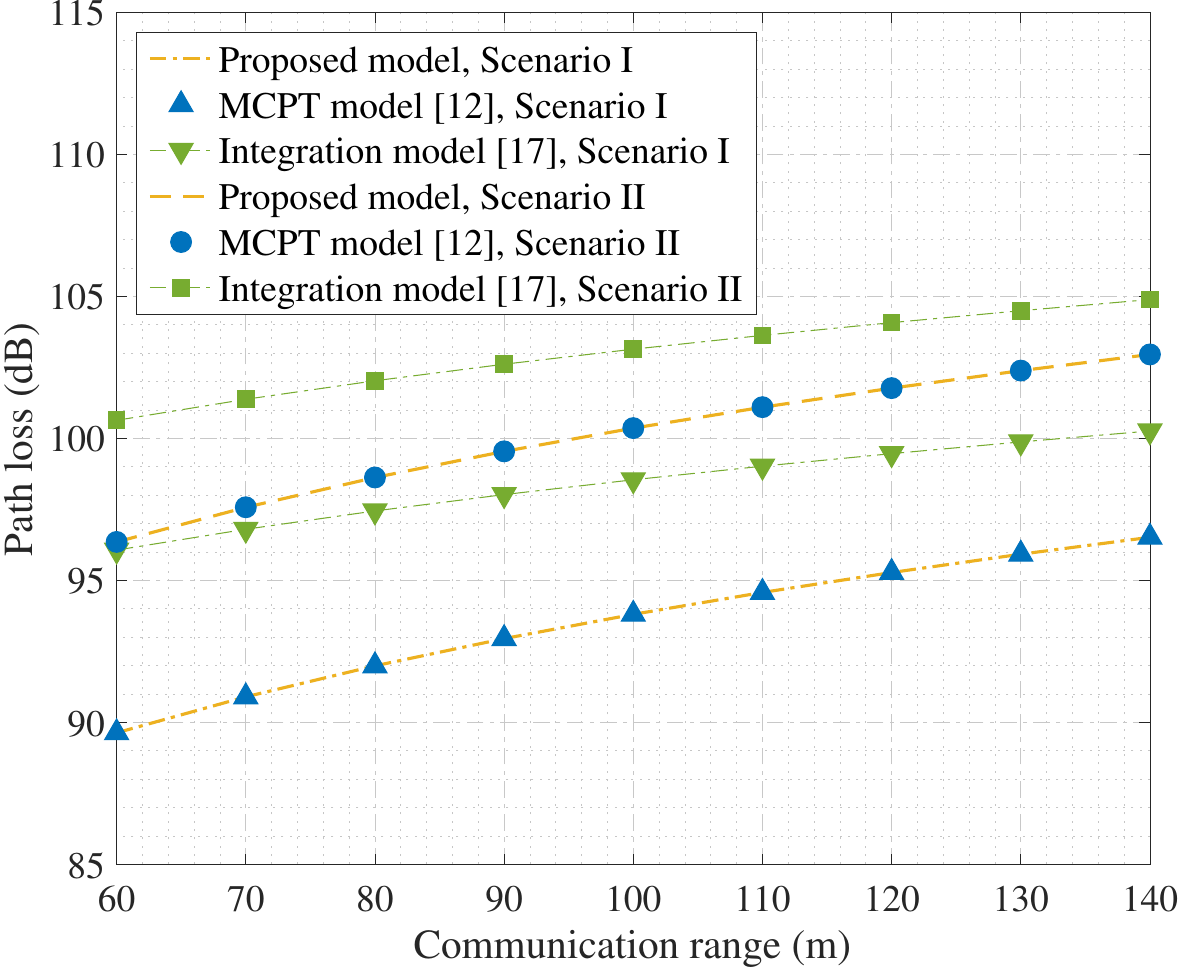}
\centering
\caption{Path loss results for the proposed model, the MCPT model \cite{ref12}, and the integration model \cite{ref17} under different scenarios.}
\label{Fig2}  
\end{figure} 
\linespread{1.5}
\begin{table}[t]
\centering
\linespread{1.0}
\caption{Parameter Settings for Validation}
\label{tab3}
\begin{tabular}{l l | l l}
\hline
{Parameters}&{Values}&{Parameters}&{Values}\\
\hline
$k^{\rm{Ray}}_s$&$0.24\,\rm{km^{-1}}$&$\beta_t$&$\pi/6$\\
$k^{\rm{Mie}}_s$&$0.25\,\rm{km^{-1}}$&$\beta_r$&$\pi/6$\\
$k_a$&$0.90\,\rm{km^{-1}}$&$\alpha_t$&$19\pi/36$\\
$\gamma$&$0.017$&$\alpha_r$&$-19\pi/36$\\
g&0.72&$A_r$&$1.92\,\rm{cm}^2$\\
$f$&0.5&$m_s$&5\\
$r_r$&0.1&$\eta$&0.5\\
\hline
\end{tabular}
\end{table}
\linespread{1.0} 

\section{Numerical Results}
In this section, we verify the proposed model by comparing it with the related MCPT model that has been validated by the outdoor experiments in two representative scenarios: i) under the canopy of roadside trees on a clear night; and ii) near~a~tall building whose facade acts as the reflective surface. During~the entire simulation, the number of collision-induced events is~set to once, because single-scattering events are dominant in short-range UV NLoS scenarios. Moreover, the number of simulated photons and their survival probability threshold are set to tens of millions and $10^{-10}$, respectively, which can ensure that the simulation error of the MCPT model is less than 0.1 dB. Then, the contributions of $\mathcal{Q}_{r,\rm{sca}}$ and $\mathcal{Q}_{r,\rm{ref}}$ to $\mathcal{Q}_r$ are investigated.

As shown in Fig.~\ref{Fig2}, obstacle parameters $s$, $w$, $\kappa$, $x_o$, and $y_o$ are set to $r/10$, $2r$, $2r$, $-3s/2$, and $r/2$, respectively, while the reflection parameters $r_r$, $m_s$, and $\eta$ of the reflection surface are set to $0.1$, $5$, and $0.5$, respectively, which are consistent with the parameter settings in~\cite{ref12}, and for the remaining parameter settings, they are summarized in Table~\ref{tab3}. From Fig.~\ref{Fig2} it can be found that the path loss curves determined by the proposed models coincide well with those obtained by the MCPT model under different scenarios, where $\vartheta_t$ and $\vartheta_r$ are set to $25^{\circ}$ and $35^{\circ}$, respectively, for Scenario I and Scenario II. Beyond that, the simulation time of the MCPT model is at least one order of magnitude longer than that of the proposed model.
\begin{figure}[t]  
\centering  
\includegraphics[scale=0.36]{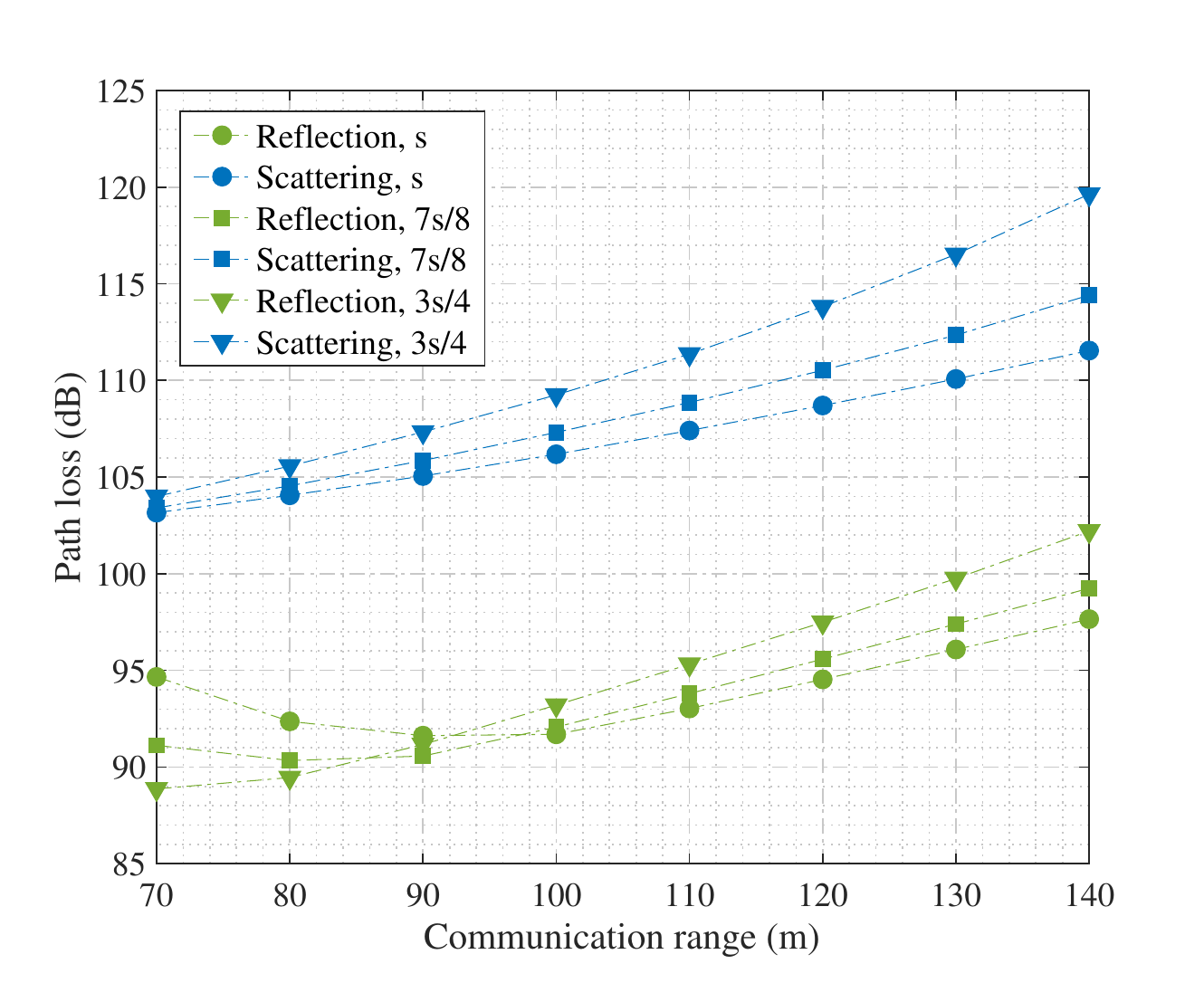}
\centering
\caption{Path loss results for scattering and reflection versus different $x_o$ values.}
\label{Fig3}  
\end{figure} 
\linespread{1.5}
\begin{table}[t]
\centering
\linespread{1.0}
\caption{Parameter Settings for Transceiver and Obstacle}
\label{tab4}
\begin{tabular}{l l | l l}
\hline
{Parameters}&{Values}&{Parameters}&{Values}\\
\hline
$\beta_t$&$\pi/12$&$w$&40 m\\
$\beta_r$&$\pi/12$&$\kappa$&80 m\\
$\alpha_t$&$2\pi/3$&$s$&30 m\\
$\alpha_r$&$-2\pi/3$&$y_o$&$r/2$\\
$\vartheta_t$&$\pi/9$&$\vartheta_r$&$\pi/9$\\
\hline
\end{tabular}
\end{table}
\linespread{1.0} 

Following that, the path loss for UV NLoS communication scenarios with and without obstacles considered is investigated under unified parameter settings. From Fig.~\ref{Fig2} it can be found that obstacle reflection can reduce the path loss of UV NLoS channels. For example, when the range is chosen as 100 m~for Scenario I, the path loss obtained by the proposed model and the integration model \cite{ref17} (no obstacles in NLoS scenarios) are 93.81 dB and 98.55 dB, respectively, which demonstrates that bypassing obstacles is not always a good option for UV NLoS communications.  

Moreover, the impact of the distance between the reflection surface and the Y-axis (i.e., $|x_o+s/2|$) on the reflected energy and the scattered energy is investigated, as presented in Fig.~\ref{Fig3}, where the model parameter settings are provided in Table~\ref{tab4}. Numerical results show that as $|x_o+s/2|$ increases, the path loss of reflected energy is overall lower than that of scattered energy. This suggests that in actual application environments, we can change the coordinates of the transceiver to adjust the value of $|x_o+s/2|$, and then adjust their elevation angles and azimuth angles to improve communication performance. 

\section{Conclusion}
In this paper, we developed a UV NLoS channel model with an obstacle, where the obstacle's thickness, width, height, and coordinates were incorporated throughout the whole modeling. To verify the proposed model, we compared it with the related MCPT model. Numerical results show that the path loss curves attained by the proposed model agree well with those obtained by the MCPT model under different cases, while its calculation time is much shorter than that of the MCPT model. Moreover, the distance between the reflection surface and the transceiver can be adjusted to improve communication performance such as data rate and bit-error rate. This work discloses that obstacle reflection can effectively reduce the channel path loss of UV NLoS systems.       

\section*{Acknowledgement}  
This work was supported by the National Key Research and Development Program of China (2023YFE0110600).


\begin{thebibliography}{1}
\bibliographystyle{IEEEtran}
\bibitem{ref1}
A.~Vavoulas, H.~G.~Sandalidis, N.~D.~Chatzidiamantis, Z.~Xu, and G.~K.~Karagiannidis,~``A survey on ultraviolet C-Band (UV-C) communications,'' \textit{IEEE Commun. Surveys Tuts.}, vol. 21, no. 3, pp. 2111--2133, 3rd Quart., 2019. 
\bibitem{ref2}
R.~Yuan and J.~Ma,~``Review of ultraviolet non-line-of-sight communication,'' \textit{Chin. Commun.}, vol. 13, no. 6, pp. 63--75, Jun. 2016. 
\bibitem{ref3}
J.~Bian, C.-X.~Wang, X.~Gao, X.~You, and M.~Zhang,~``A general 3D non-stationary wireless channel model for 5G and beyond,''~\textit{IEEE Trans. Wireless Commun.}, vol. 20, no. 5, pp. 3211--3224, May 2021. 
\bibitem{ref4}
C.~Chen, D.~A.~Basnayaka, A.~A.~Purwita, X.~Wu, and H.~Haas,~``Wireless infrared-based LiFi uplink transmission with link blockage and random device orientation,'' \textit{IEEE Trans. Commun.}, vol. 69, no. 2, pp. 1175--1188, Feb. 2021.
\bibitem{ref5}
L.~Feng, H.~Yang, R.~Q.~Hu, and J.~Wang,~``MmWave and VLC-based indoor channel models in 5G wireless networks,'' \textit{IEEE Wirel. Commun.}, vol. 25, no. 5, pp. 70--77, Oct. 2018.
\bibitem{ref6}
G.~Chen, T.~Wu, F.~Yang, T.~Wang, J.~Song, and Z.~Han,~``Ultraviolet-based UAV swarm communications: potentials and challenges,'' \textit{IEEE Wirel. Commun.}, vol. 29, no. 5, pp. 84--90, Oct. 2022.
\bibitem{ref7}
T.~Wu, T.~Cao, F.~Yang, J.~Song, J.~Cheng, and Z.~Han,~``Ultraviolet-based indoor wireless communications: potentials, scenarios, and trends,'' \textit{IEEE Commun. Mag.}, vol. 62, no. 3, pp. 82--88, Dec. 2023. 
\bibitem{ref8}
S.~Wang, R.~Yuan, M.~Peng, Z.~Wang, X.~Chu, S.~Di, K.~Sun, and D.~Zhang,~``Non-line-of-sight ultraviolet positioning using two photon-counting receivers,''~\textit{GLOBECOM}, Kuala Lumpur, Malaysia, Dec. 2023, pp. 3682-3687.
\bibitem{ref9}
T.~Wu, F.~Yang, J.~Song, J.~Ma, and P.~Su,~``Modeling of UV diffused-LOS communication channel incorporating obstacle and its applicability analysis,'' \textit{Opt. Lett.}, vol. 46, no. 18, pp. 4578--4581, Sep. 2021.
\bibitem{ref10}
T.~Wu, T.~Cao, F.~Yang, J.~Song, J.~Cheng, and Z.~Han,~``Modeling of UV diffused-LoS communication channel incorporating obstacles: an integration perspective,''~\textit{IEEE Trans. Wireless Commun.}, \textit{Early Access}, May 2024.
\bibitem{ref11}
H.~Zhang, H.~Yin, H.~Jia, J.~Yang, and S.~Chang,~``Study of effects of obstacle on non-line-of-sight ultraviolet communication links,'' \textit{Opt. Express}, vol. 19, no. 22, pp. 21216--21226, Oct. 2011.
\bibitem{ref12}
T.~Cao, X.~Gao, T.~Wu, C.~Pan, and J.~Song,~``Reflection-assisted non-line-of-sight ultraviolet communications,''~\emph{J. Lightwave Technol.}, vol. 40, no. 7, pp. 1953--1961, Apr. 2022. 
\bibitem{ref13}
M.~R.~Luettgen, J.~H.~Shapiro, and D.~M.~Reilly,~``Non-line-of-sight single-scatter propagation model,'' \textit{J. Opt. Soc. Amer. A, Opt. Image Sci.}, vol. 8, no. 12, pp. 1964--1972, Dec. 1991.
\bibitem{ref14}
T.~Wu, J.~Ma, P.~Su, R.~Yuan, and J.~Cheng,~``Modeling of short-range ultraviolet communication channel based on spherical coordinate system,'' \textit{IEEE Commun. Lett.}, vol. 23, no. 2, pp. 242--245, Feb. 2019.
\bibitem{ref15}
Z.~Xu, H.~Ding, B.~M.~Sadler, and G.~Chen,~``Analytical performance study of solar blind non-line-of-sight ultraviolet short-range communication links,'' \textit{Opt. Lett.}, vol. 33, no. 16, pp. 1860--1862, Aug. 2008.
\bibitem{ref16}
T.~Cao, T.~Wu, C.~Pan, and J.~Song,~``Single-collision-induced path loss model of reflection-assisted non-line-of-sight ultraviolet communications,'' \textit{Opt. Express}, vol. 30, no. 9, pp. 15227--15237, Apr. 2022. 
\bibitem{ref17}
Y.~Zuo, H.~Xiao, J.~Wu, Y.~Li, and J.~Lin,~``A single-scatter path loss model for non-line-of-sight ultraviolet channels,''~\textit{Opt. Express}, vol. 20, no. 9, pp. 10359--10369, Apr. 2012.
\end{thebibliography}
\end{document}